\documentclass[prl,twocolumn,showpacs]{revtex4}

\usepackage{graphicx}
\usepackage{amsmath}
\usepackage{amssymb}

\begin{document}
\def\a{\alpha}
\def\b{\beta}
\def\e{\varepsilon}
\def\d{\delta}
\def\k{\kappa}
\def\l{\lambda}
\def\m{\mu}
\def\t{\tau}
\def\n{\nu}
\def\o{\omega}
\def\r{\rho}
\def\s{\sigma}
\def\S{\Sigma}
\def\G{\Gamma}
\def\D{\Delta}
\def\O{\Omega}

\def\ra{\rightarrow}
\def\ua{\uparrow}
\def\da{\downarrow}
\def\pd{\partial}
\def\bk{{\bf k}}
\def\br{{\bf r}}
\def\bq{{\bf q}}
\def\bn{{\bf n}}

\def\be{\begin{equation}}
\def\ee{\end{equation}}
\def\bea{\begin{eqnarray}}
\def\eea{\end{eqnarray}}
\def\nn{\nonumber}
\def\lb{\label}
\def\pref#1{(\ref{#1})}

\title{Impurity clusters and localization of nodal quasiparticles
in \emph{d}-wave superconductors}

\author{Yu.G. Pogorelov}

\affiliation{IFIMUP/Departamento de F\'{\i}sica, Universidade do
Porto, R. Campo Alegre, 687, Porto, 4169-007, Portugal}

\author{M.C. Santos}

\affiliation{Departamento de F\'{\i}sica, Universidade de Coimbra,
R. Larga, Coimbra, 3004-535, Portugal}

\author{and V.M. Loktev}

\affiliation{Bogolyubov Institute for Theoretical Physics, 14b
Metrologichna str., 03143 Kiev, Ukraine}

\begin{abstract}
The long disputed issue of the limiting value of quasiparticle
density of states $\r(0) = \r ( \e \to 0 )$ in a \emph{d-}wave
superconductor with impurities (\emph{vs} its linear vanishing,
$\r_0(\e) \propto |\e|/\D$, near the nodal point $\e = 0$ in a pure
system with the gap parameter $\D$) is discussed. Using the
technique of group expansions of Green functions in complexes of
interacting impurities, it is shown that finite $\r(0)$ value is
possible if the (finite) impurity perturbation $V$ is spin-dependent
(magnetic). The found value has a power law dependence on the
impurity concentration $c$: $\r(0) \propto \r_N c^{n}$, where $\r_N$
is the normal metal density of states and $n \geq 2$ is the least
number of impurities in a complex that can localize nodal
quasiparticle. This result essentially differs from the known
predictions of self-consistent approximation: $\r(0) \propto \r_N
\sqrt {c/\r_N\D}$ (for the unitary limit $V \to \infty$) or $\r(0)
\propto (\D/cV^2)\exp(-\D/cV^2\r_N)$ (for the Born limit $|V|\r_N
\ll 1$). We predict also existence of a narrow interval (mobility
gap) around the Fermi energy, where all the states are localized on
proper impurity clusters, leading to exponential suppression of
low-temperature kinetics.
\end{abstract}
 \maketitle

\section{Introduction}
Key objects for the low-temperature physics of high-$T_{c}$
superconductors with planar lattice structure and \emph{d}-wave
superconducting (SC) order are quasiparticles of low energy, $\e \ll
\D \ll \e_{\rm F}$ (where $\D$ is the SC gap parameter and $\e_{\rm
F}$ is the Fermi energy). In a clean \emph{d}-wave system, they have
``conical'' energy dispersion: $\e_{\bk} \approx \hbar \sqrt{v_{\rm
F}^2 q_1^2 + v_{\D}^2 q_2^2}$, if the wave vector $\bk$ is close to
the nodal points $\bk_j$ ($|\bk_j| = k_{\rm F},\,j=1,\dots,4$, Fig.
\ref{fig1}) on the Fermi surface. Here $q_1$ and $q_2$ are the
radial and tangential components of the relative quasimomentum $\bq
= \bk  - \bk_j$  and the ``gap'' velocity $v_\D$ is small beside the
common Fermi velocity $v_{\rm F}$: $v_{\D}/v_{\rm F} = \D/\e_{\rm F}
= \b \ll 1$. This defines the linear energy dependence of
quasiparticle density of states (DOS): $\r_0(\e) \approx
\r_N|\e|/\D$ ($\r_N$ is the normal metal DOS), and respective power
laws in temperature for thermodynamical quantities, in agreement
with experimental data \cite{tsuei}.

\begin{figure}
    \includegraphics[width=6.5cm]{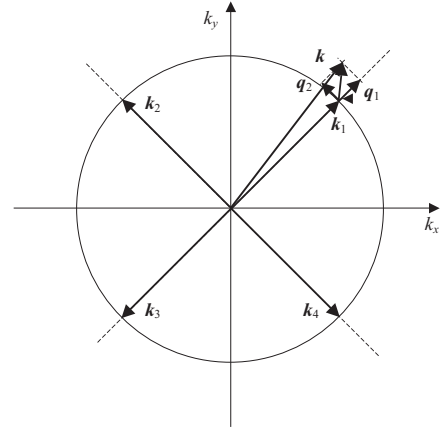}\\
  \caption{Schematic of local coordinates near nodal points
  in the Brillouin zone of a \emph{d}-wave superconductor.}
  \lb{fig1}
\end{figure}

But this ``clean'' picture can fail at the lowest temperatures,
dominated by the lowest energy excitations. By many analogies in the
condensed matter, such excitations are expected to be the most
sensitive to presence of impurities (almost inevitable in
high-$T_{c}$ compounds). In particular, impurities can produce low
energy resonances in the \emph{d}-wave materials \cite{bal,pog}, and
even the possibility for impurity localization of quasiparticles was
discussed \cite{bal,lee}. Traditionally, such effects were treated
using certain forms of single-impurity approximation, as T-matrix
approximation \cite{pet-pin} or its generalization to the
self-consistent T-matrix approximation (SCTMA)
\cite{sch-rink,hirsch}. Then, within the SCTMA scope, a finite
limiting value of DOS, $\rho(0) = \lim_{\e\to 0}\rho(\e)$, was
predicted either in the Born limit, $|V|\rho_N \ll 1$ \cite{gork},
and in the unitary limit, $|V|\rho_N \gg 1$ \cite{sch-rink,lee}, for
impurity perturbation $V$. However it was shown later that, for a
properly defined SCTMA solution, $\rho(0)$ should vanish (at any
finite $V$) \cite{lp2}. The issue of finite or zero limit for DOS is
of crucial importance for low temperature physics of real
high-$T_{c}$ systems. Thus, supposing existence of $\rho(0)\neq 0$,
P.A. Lee arrived at the prediction that the quasiparticle electric
a.c. conductivity $\s(\o)$ in the static limit $\o \to 0$ should
tend to a \emph{universal} value $\s(0) = (e^2/\pi^2\hbar\b)$
\cite{lee}, independent of impurity parameters at all! Similarly, a
universal limit was predicted for the ratio of thermal conductivity
to temperature, $\left(\k/T\right)_{T\to 0} \to [n k_{\rm
B}^2/(3\hbar v_{\rm F} v_{\D}d)](\b^{-1} + \b)$ \cite{graf,durst}
(where $n$ is the number of conducting planes per unit cell and $d$
the distance between them). But no such behavior would be possible
for the above referred alternative of vanishing DOS at $\e\to 0$.
This alternative was predicted in various analytical forms:
sublinear $\r(\e) \propto |\e|^\zeta$ with $\zeta < 1$ \cite{ners}
(also concluded from numerical solution of Bogolyubov-de Gennes
equations \cite{atk}), linear $\r(\e) \propto |\e|$ (for
non-magnetic impurities) \cite{senth}, or superlinear $\r(\e)
\propto \e^2$ (for magnetic impurities) \cite{senth} and $\r(\e)
\propto |\e|/\ln^2(\D/|\e|)$ \cite{lp}. It should be also mentioned
that some field theory approaches to this problem even resulted in
diverging DOS: $\r(\e) \propto 1/\left[|\e|\ln^2(\D/\e)\right]$
\cite{pepin} or $\r(\e) \propto \exp\left[A\sqrt{\ln
(\D/\e)}\right]/|\e|$ \cite{fabr}. Up to the moment, the
experimental checks for universal behavior in the low-frequency
electric transport \cite{hardy,basov} and low-temperature heat
transport \cite{tail,suther} do not provided a definitive support
for any of these scenarios.

It is important to notice that either the SCTMA approach of Refs.
\cite{lee,durst,lp} and the field-theory models of Refs.
\cite{ners,senth,fabr} are only justified if the relevant
quasiparticle states are extended (non-localized). This requires
that a quasiparticle with wavevector $\bk$ have the mean free path
$\ell$ longer than its wavelength $\l = 2\pi/k$, as expressed by the
known Ioffe-Regel-Mott (IRM) criterion \cite{irm}. Also the
derivation of the above mentioned universal limits fully relies on
the Kubo formula integrals over the \emph{extended} states
(involving a certain lifetime $\t = \ell/v$ for given quasiparticle
velocity $v$). But in presence of \emph{localized} states, the
concepts of mean free path and lifetime do not make sense for them,
and the Kubo formula should be reconsidered for the corresponding
range of the spectrum, e.g., as for hopping transport in
semiconductors \cite{zhang}. Fortunately, the analysis of DOS can be
done independently from the controversial issue of universal
conductivity (postponing it for more detailed future treatment).

As to the limiting DOS values, it was shown \cite{lp1} that the
SCTMA approach in fact admits two solutions. One of them, called
SCTMA-1, leads to finite $\r(0) \neq 0$ as given in Refs.
\cite{lee,gork}, but the other solution, SCTMA-2, yields in
vanishing $\r(\e \to 0) \to |\e|/\ln^2(\D/|\e|)$. And only SCTMA-2
proves to satisfy the IRM criterion at $\e \to 0$, thus appearing as
the only valid SCTMA solution in this limit. Applying the same check
to the power law solutions $\r(\e) \propto |\e|^\zeta$, one finds
them to satisfy the IRM criterion only if $\zeta > 1$, which is not
actually the case in Refs. \cite{ners, atk} and \cite{senth} (for
non-magnetic impurities). Thus, it could be thought that the
ambiguity is resolved in favor of the superlinearly vanishing
SCTMA-2 DOS.

However the SCTMA (or field-theoretical) analysis can not be
considered fully comprehensive for the real quasiparticle spectrum
in a disordered system, if the localized states are also admitted. A
more complete description is possible, noting that the SCTMA
self-energy is only the first term of the so-called group expansion
(GE) \cite{lp}, where this term describes all the processes of
quasiparticle scattering on \emph{isolated} impurities. In
accordance with the above IRM check, such processes can not produce
localized states near zero energy. But if the SCTMA contribution to
DOS vanishes in this limit, the importance can pass to the next
terms of GE, related to scattering (and possible localization) of
quasiparticles on random groups (clusters) of impurities. The
essential point is that their contribution to DOS is mostly defined
by the \emph{real} parts of Green functions which do not need
self-consistency corrections and thus remain valid for the energy
range of localized states (where the IRM criterion does not hold).

The known approaches to impurity cluster effects in disordered
\emph{d}-wave superconductors, either numerical \cite{zhu1,zhu2} and
analytical \cite{pepin}, were contradictory about DOS and did not
conclude definitely on localization. A practical analysis of such
effects within the GE framework for an \emph{s}-wave superconductor
was proposed recently \cite{lp2}, using a special algebraic
technique in the limit $\e \to \D$. The present work develops a
similar technique for a disordered \emph{d}-wave system in the limit
$\e \to 0$. We shall consider two main types of impurities,
non-magnetic (NM) and magnetic (M), and show that only M-impurities
can provide a finite (but different from the SCTMA-1 value) DOS,
replacing the SCTMA-2 solution in a narrow vicinity of the (shifted)
nodal energy and manifesting onset of localization there. The latter
should produce, instead of universal conductivity, its exponential
suppression at sufficiently low temperatures.

\section{Formulation of problem}
In more detail, we use the low energy Hamiltonian

\bea
 H & = & \hbar \sum_\bk \psi_\bk^\dagger \left(v_{\rm F}q_1 \hat\t_3 +
 v_\D q_2 \hat\t_1\right) \psi_\bk \nn \\
 & & \qquad\qquad + \frac {1} {N} \sum_{{\bf p},\bk,\bk^\prime}
 {\rm e}^{i(\bk - \bk^\prime)\cdot{\bf p}} \psi_\bk^\dagger \hat V
 \psi_{\bk^\prime}
 \lb{eq1}
\eea

\noindent where $\bk$ are restricted to vicinities of nodal points
$\bk_j$, the Nambu spinors $\psi_\bk^\dagger = \left( a_\bk^\dagger,
\a_{-\bk}\right)$ are composed of normal metal Fermi operators, $N$
is the number of lattice sites and $\hat\t_i$ are Pauli matrices.
The second term in Eq. \ref{eq1} corresponds to the so called
Lifshitz impurity model. It describes quasiparticle scattering on
random sites $\bf p$ with concentration $c = \sum_{\bf p} 1/N \ll 1$
in lattice of $N$ sites, and the perturbation matrix is:

\be \hat V = V \left\{\begin{array}{c}
                 \hat\t_3,\quad{\rm NM}, \\
                 1,\qquad{\rm M},
                                \end{array}\right.
                                \lb{eq2}
                                \ee

\noindent  for specific types of impurities. We calculate the
Fourier transform of retarded Green function (GF) matrix

\be
 \left\langle\left\langle \psi_\bk| \psi_{ \bk^\prime}^\dagger
 \right\rangle\right\rangle_\e = i\int_0^\infty {\rm e}^{i(\e +
 i0)t} \left\langle \left\{\psi_\bk,\psi_{\bk^\prime}^\dagger
 \right\} \right\rangle dt
 \lb{eq3}
\ee

\noindent where $\langle\dots\rangle$ is the quantum-statistical
average with the Hamiltonian, Eq. \ref{eq1}, and $\{a,b\} = ab + ba$
is the anticommutator. The most important momentum-diagonal GF is
presented as

\be \left\langle\left\langle \psi_\bk| \psi_{ \bk}^\dagger
 \right\rangle\right\rangle_\e = (\hat G_\bk^{-1} -
 \hat\S_\bk )^{-1},
\lb{eq4}
 \ee

\noindent where $\hat G_\bk(\e) = \left[\e + \hbar\left(v_{\rm
F}q_1\hat\t_3 + v_\D q_2\hat\t_1\right)\right]/\left(\e^2 -
\e_\bk^2\right)$ is the GF matrix of a clean system and the
self-energy matrix $\hat \S_\bk$ is expanded into the GE series
\cite{pog}:

\bea
 \hat\S_\bk &=& c \hat T \left[1 + c \sum_{\bf n \neq 0}\left(\hat
A_\bn \cos\bk \cdot \bn + \hat A_\bn^2\right) \right.  \nn\\
& & \qquad \qquad \qquad\qquad \left.\times \left(1 - \hat A_\bn^2
\right)^{-1} + \dots \right].
 \lb{eq5}
  \eea

\noindent Here the T-matrix $\hat T(\e) = \hat V \left(1 - \hat G
\hat V \right)^{-1}$, with local GF matrix $\hat G(\e) =
N^{-1}\sum_\bk \hat G_\bk$, describes all the scattering processes
on a single impurity center. It can have resonance behavior near a
certain energy $\e_{res} < \D$, such that ${\rm Re\,\det} \left[1 -
\hat G\left(\e_{res}\right) \hat V \right] = 0$ \cite{bal,pog,lp2},
however at $\e \to 0$ it simply tends to a certain real matrix
$\hat{T}(0)$, to be specified below.

The next to unity term in the brackets of Eq. \ref{eq5} describes
scattering on pairs of impurities, separated by all possible lattice
vectors $\bn$, while the dropped terms are for impurity triples and
so on. The building block of all GE terms is the matrix $\hat
A_\bn(\e) = N^{-1} \sum_\bk {\rm e}^{i\bk\cdot\bn}\hat G_\bk \hat T$
which represents the effective (energy dependent) interaction
between two impurities at given separation $\bn$. Its zero energy
limit is

\be
 \hat{A}_\bn(0) = \left(g_\bn\hat\t_3 +
 f_\bn\hat\t_1\right)\hat T(0),
 \lb{eq6}
 \ee

\noindent with real functions $g_\bn = -(\hbar v_{\rm F}/N)\sum_\bk
{\rm e}^{i\bk\cdot\bn}q_1/\e_\bk^2$ and $f_\bn = -(\hbar v_\D/N)
\sum_\bk {\rm e}^{i\bk\cdot\bn}q_2/\e_\bk^2$. Then the DOS reads:

\be
 \r(\e) = \frac 1 {N\pi} \sum_\bk {\rm Im \,Tr} \left\langle
\left\langle \psi_\bk| \psi_{ \bk}^\dagger \right\rangle
\right\rangle_\e,
 \lb{eq7}
 \ee

\noindent and in absence of impurities it is $\r_0(\e) =
\pi^{-1}{\rm Im} g(\e) $, where $g(\e) = N^{-1}{\rm Tr} \sum_\bk
\hat G_\bk \approx (2\r_N \e/\D) \ln (2i\D/\e)$ \cite{lp}, vanishing
linearly with $\e \to 0$. In presence of impurities, Eqs. \ref{eq4}
and \ref{eq5} generally relate the limit $\r(\e \to 0)$ with the
imaginary and traceful part of the self-energy matrix: $\gamma =
\frac 1 2 {\rm Tr \,Im} \hat \S_\bk(0)$. If the main contribution to
$\gamma$ comes from the GE pair term, where the share of $\hat
A_\bn\cos\bk\cdot\bn$ is negligible beside that of $\hat A_\bn^2$,
$\gamma$ can be considered momentum-independent. In this
approximation, supposing also $\gamma \ll \D$, we obtain

\be
 \r(0) \approx \r_N \frac {2\gamma} {\pi\D} \ln \frac \D \gamma \ll \r_N,
 \lb{eq7a}
 \ee

\noindent and the following task is reduced to proper calculation of
$\gamma$ in function of the impurity perturbation parameters.

Since the above mentioned matrices $\hat T(0)$ and $\hat A_\bn$ are
real, the imaginary part of the GE pair term is generated by the
poles of $\left(1 - \hat A_\bn^2 \right)^{-1}$. If there is no such
poles, one has to search for contributions to $\gamma$ from next
order GE terms. In this course, for an \emph{l}th order GE term, the
imaginary part is related to the poles of the inverse of a certain
\emph{l}th degree polynomial in $\hat A_\bn,\dots,\hat
A_{\bn_{l(l-1)/2}}$ (where $\bn_1,\dots,\bn_{l(l-1)/2}$ are all
possible separations between \emph{l} impurities).

Generally, in the energy spectrum of a crystal with impurities, one
can distinguish certain intervals where DOS is dominated by
contributions from band-like states, single impurities, impurity
pairs, triples, etc. \cite{lif}. Then, for instance, in the
pair-dominated energy interval, each discrete peak (by an impurity
pair at given separation $\bn$ in the lattice) experiences small
shifts, due to the effects by neighbor impurities of such a pair,
different in different parts of the system. These shifts produce a
broadening of pair peaks, and if it is wider than the distance
between the peaks, the resulting continuous pair-dominated spectrum
can be effectively described, passing from summation in discrete
$\bn \neq 0$ in Eq. \ref{eq5} to integration in continuous $\br$
(for $r > r_0$ where $r_0 \sim a$). Such possibility was shown long
ago for normal electron spectrum \cite{iv}, and it is even more
pronounced for the superconducting system where the pair
contribution to DOS at a given energy can come from multiple pair
configurations (see Figs. \ref{fig2},\ref{fig4} below) and this
multiplicity turns yet much greater at involving neighbor impurities
into each configuration. An approach to analogous problem in
\emph{s}-wave superconductors was proposed recently \cite{lp2},
using an algebraic isomorphism of the matrices of interaction
between impurities in that system to common complex numbers.

\section{Non-magnetic impurities}
Remarkably, the same isomorphism is also found for matrices of zero
energy interaction $\hat A_\bn$ between NM impurities in the
\emph{d}-wave system. In this case, we have explicitly: $\hat T(0) =
\tilde V \hat \t_3 $, where $\tilde V = V/(1 - V g_{as})$ and
$g_{as} \approx \r_N \ln|1/\r_N\e_{\rm F}-1|$ is the factor of
particle-hole asymmetry (away from half-filling: $\r_N\e_{\rm F}
\neq 1/2$). Then the interaction matrices, Eq. \ref{eq6}, are
presented as $\hat A_\bn =\tilde V g_\bn + i \tilde V f_\bn
\hat\t_2$, thus pertaining to the general two-parametric family
$\hat C (x,y) = x + i y \hat{\tau}_{2}$ with real $x,y$. This family
forms an algebra with the product $\hat C(x,y) \hat C
(x^\prime,y^\prime) = \hat C (xx^\prime - yy^\prime,yx^\prime +
xy^\prime)$, isomorphic to that considered in Ref. \cite{lp2} and to
the algebra $\mathbb{C}$ of common complex numbers: $(x +
iy)(x^\prime + iy^\prime) = xx^\prime - yy^\prime + i(yx^\prime +
xy^\prime)$. By this isomorphism, the real matrix $\hat A_\bn$ is
related to a ``complex number'' $A_\bn = \tilde V g_\bn + \hat i
\tilde V f_\bn$, where the ``imaginary unity'' $\hat i$ corresponds
to the \emph{real} matrix $\hat i \equiv i \hat \t_2$. Using such
``complex'' representation and the above mentioned passage from
summation in $\bn$ to integration in $\br$, we can write the pair
contribution to $\gamma$ in the form:

\begin{figure}
  \includegraphics[width=9cm]{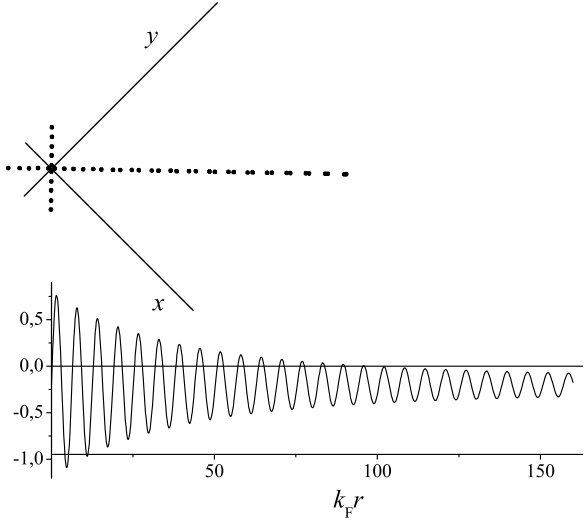}\\
  \caption{The poles of the integrand in Eq. \ref{eq8} in function of
  pair separation vector $\br = (x,y)$, are located along the nodal
  the direct space), where one of the pole conditions, $f_\br = 0$, holds
  identically. Another pole condition, $|g_\br|  = 1/|\tilde V |= |g_{as} - 1/V|$,
  is reached at discrete points (here at the  choice of parameters
  $\b =  0.05, \tilde V  g_{as} = 5$). The plot below shows the related
  behavior of $g_\br - 1/\tilde V$ for $\br$ along the nodal direction.}
  \lb{fig2}
\end{figure}

\be
 c^2 {\rm Im} \int_{r > r_0} \frac {d\br}{a^2}  \Re \frac{p(\br) + \hat i q(\br)}
 {s(\br) + \hat i t(\br)},
 \lb{eq8}
 \ee

\noindent with $s = 1 - \tilde V^2(g_\br^2 - f_\br^2), \, t = - 2V^2
g_\br f_\br$ and some continuous functions $p(\br),q(\br)$. Here the
symbol $\Re$ means the ``real'' (traceful) part of the ``complex''
integrand, while the common imaginary part ${\rm Im}$ is produced by
its poles. These are attained at such separations $\br = \br^\ast$
of an impurity pair that $s(\br^\ast) = t(\br^\ast) = 0$, which
requires $|g_{\br^\ast}| = \tilde V^{-1}$ and $f_{\br^\ast} = 0$.
Direct calculation of $g_\br$ and $f_\br$ shows that all possible
$\br^\ast$ lie on nodal directions, forming identical finite series
along them (like those in Fig. \ref{fig2}). Also we note that poles
only exist at high enough band filling, $\r_N\e_{\rm F} \geq (1 +
\rm e)^{-1}$ and strong enough perturbation parameter, $V \geq (\r_N
- g_{as})^{-1}$, as chosen in Fig. \ref{fig2}. If so, it is suitable
to pass in the vicinity of each $\br^\ast$ from integration in the
components $r_1,r_2$ of vector $\br$ to that in the components $s,t$
of the ``complex'' denominator:

\be
  {\rm Im} \sum_{\br^\ast} \int ds dt
{\mathcal{J}_{\br^\ast}(s,t)}\frac{p(s,t)s + q(s,t)t}
 {s^2 + t^2},
 \lb{eq9}
 \ee

\noindent and for any $\br^\ast$ the transformation Jacobian

\[{\mathcal J}_{\br^\ast}(s,t) = \left.\frac{\pd\left(r_1,r_2\right)}
{\pd \left(s,t\right)}\right|_{\br = \br^\ast}\]

\noindent is real and non-singular. Then the singularity in the
denominator of Eq. \ref{eq9} is canceled by the vanishing residue in
the numerator and, even at formal existence of poles in Eq.
\ref{eq8}, they give no contribution to the zero energy DOS.
Mathematically, this simply follows from an extra dimension at 2D
integration, giving zero weight to the isolated poles.

The above conclusion can be immediately generalized for any $l$th
order GE term ($l\geq 3$), where the integrand is again presented as
$(p + \hat i q)/(s + \hat i t)$ and $p, q, s, t$ are now continuous
functions of $N_l = 2(l - 1)$ independent variables (components of
the vectors $\br_1, \dots, \br_{l-1}$) in the configurational space
$\mathcal S_l$. This integrand can have simple poles on some $(N_l -
2)$-dimensional surface $\mathcal A_l$ in $\mathcal S_l$ (under
easier conditions than for $l = 2$). Then the $N_l$-fold integration
can be done over certain coordinates $u_1,\dots,u_{N_l-2}$ in
$\mathcal A_l$ and over the components $s,t$ of the ``complex''
denominator in the normal plane to $\mathcal A_l$:

\bea &&{\rm Im} \int du_{1}\dots du_{N_{l}-2} {\mathcal J}\left(u_1,
 \dots,u_{N_l-2}\right)\nn\\
 &\times & \int ds dt\frac{ p\left(
 \dots,s,t\right)s +
 q\left(
 \dots,s,t\right)t}{s^2+t^2},
 \lb{eq10}
 \eea

\noindent with a non-singular Jacobian $\mathcal J$. The latter
integral has no imaginary part by the same reasons as for Eq.
\ref{eq9}. Thus, it can be concluded that perturbation by
NM-impurities in a \emph{d}-wave system can not produce localized
quasiparticles of zero energy, and this directly follows from the
indicated isomorphism of the interaction matrices to the algebra
$\mathbb C$ of complex numbers.

Moreover, the same conclusion is also valid for yet another type of
NM-perturbation, due to locally perturbed SC order by the matrix
$\hat V = V \hat\t_1$. In this case, the interaction matrix:

 \[\hat A_\bn  =  \frac {V}{1-V^2 g_{as}^2}\left[ f_\bn + V g_{as}g_\bn
 +  i \left(g_\bn - V g_{as}f_\bn \right) \hat\t_2 \right],\]

\noindent pertains to the same family $\hat C(x,y)$ as in the above
case, hence leading to the same absence of contribution to the zero
energy DOS.

\section{Magnetic impurities}
However, an essential cluster contribution to zero energy DOS can be
produced by M impurities with the scalar (see Eq. \ref{eq2}) local
perturbation $\hat V = V $, the respective T-matrix in the zero
energy limit being:

\be
 \hat T(0) =  V  (1 -  V  \hat G)^{-1} = V \frac {1  +  V g_{as}
 \hat\t_3}{1   -  V^2 g_{as}^2}.
 \lb{eq11}
\ee

\begin{figure}
\center\includegraphics[width=8cm]{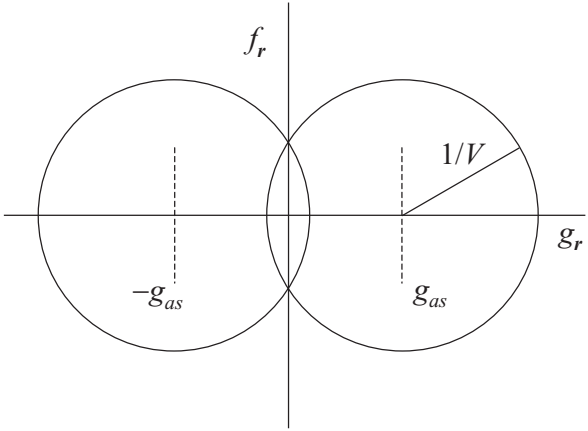}
 \caption{\lb{fig3}
 Circular trajectories in the space of variables $g_\br,f_\br$,
 corresponding to the poles of GE denominator, Eq. \ref{eq13}, for M
 impurities at the choice of $V = 1/\r_N$ and $g_{as} = 0.84 \r_N$.}
 \end{figure}

\noindent Unlike the traceless $\hat T(0)$ for NM-impurities, this
matrix produces a finite shift of the nodal point itself, from $\e =
0$ to $\e = \e_0 = cV(1 - V^2 g_{as}^2)^{-1}$. As usual, this can be
absorbed into the Fermi level position by shifting all the energy
arguments of considered GF's, then DOS in T-matrix (or SCTMA)
approximation will vanish at $\e \to \e_0$ (in the same way as at
$\e \to 0$ for NM-impurities), fixing the relevant limit in presence
of M impurities. The following treatment of higher order GE terms
involves the matrix of interaction between M-impurities:

\be
 \hat{A}_\br  =  V \frac{(\hat\t_3 + Vg_{as})g_\br +
( \hat\t_1 - i  V g_{as}\hat\t_2) f_\br} {1   -  V^2 g_{as}^2}.
 \lb{eq12}
  \ee

\noindent  Notably, it does not fit the $\mathbb C$ algebra, and
though being harder technically, this permits to expect that
M-impurities can effectively contribute to the zero energy DOS. In
fact, the straightforward calculation of the corresponding GE pair
term leads to the general matrix expression

\be
 (1 - \hat A_\br^2)^{-1}  = \frac {N_0 + N_1\hat\t_1 + i
  N_2\hat\t_2 +N_3\hat\t_3}
  {D_\br }.
 \lb{eq13}
 \ee

\noindent Here $N_j$'s in the numerator are certain functions of
$g_\br, f_\br$ and the denominator $D$ is the 4th grade polynomial:

\[ D  =  \left[f_\br^2 + \left(g_\br - g_{as}\right)^2 - \frac 1 {V^2}\right]
 \left[f_\br^2 + \left(g_\br + g_{as}\right)^2 - \frac 1 {V^2}\right]. \]

\noindent Zeroes of $D$ in the space of variables $g_\br,f_\br$ form
two circular trajectories of radius $1/V$ centered at $\pm g_{as}$,
as shown in Fig. \ref{fig3} (cf. to their location in the two outer
points of these circles, $f_\br = 0,\: \pm g_\br = g_{as} + 1/V$, in
the NM case). It is this extension of singularities, from isolated
points to continuous trajectories, that allows finite imaginary part
of the 2D integral

\be
 \gamma \approx \frac{c^2 V}{1 - V^2 g_{as}^2} {\rm Im}
 \int \frac{d\br}{a^2} \frac{N_0 + V g_{as}N_3} {D_\br}.
 \lb{eq14}
 \ee

\begin{figure}
\center\includegraphics[width=9cm]{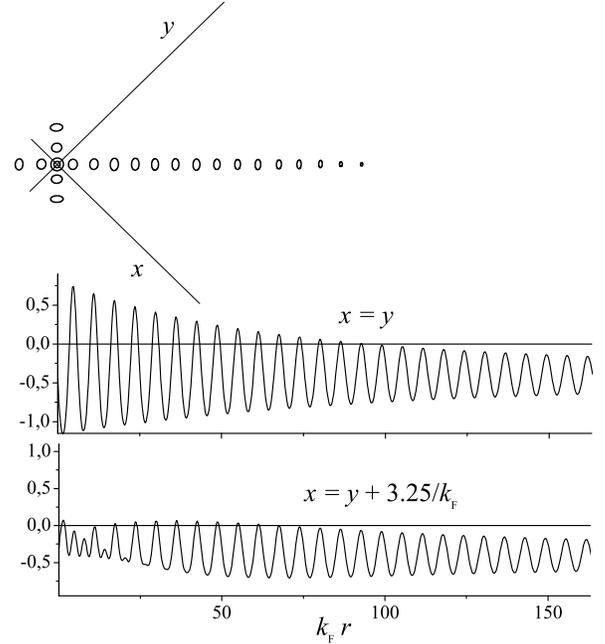}
 \caption{\lb{fig4}
 In the $\br$-plane, the trajectories of zeroes of $D_\br$  form
 continuous loops within the stripes of $\sim v_{\rm
 F}/(v_\D|\d|k_{\rm F})$ length and $\sim 1/(|\d|k_{\rm F})$ width
 along the nodal axes. The plots below present the pair GE term
 denominator $D_\br$ (at the same choice of parameters as in Fig.
 \ref{fig3}), seen along the stripe axis and along the stripe
 border.}
 \end{figure}

\noindent On the other hand, it is assured by the fact that the
traceful numerator $N_0 + V g_{as}N_3$ in Eq. \ref{eq14} does not
vanish on these trajectories. Quantitative analysis is simplified in
the case of $Vg_{as} \approx 1$, when the small parameter $\d = 1 -
V^2 g_{as}^2$ (such that $|\d| \ll 1$) defines a low energy
resonance at $\e_{res} \approx \D\d/[V\r_N \ln(1/\d^2)]$ \cite{lp3}.
In this case we have simply $N_0 + V g_{as}N_3 \approx -2
g_\br^2/V^2$. Direct numeric calculation of the functions
$g_\br,f_\br$ shows that the trajectories $D = 0$, when presented in
variables $r_{1,2}$, form multiple loops (seen in the inset of Fig.
\ref{fig4}) of total number $\sim v_{\rm F}/(v_\D|\d|)$, each
contributing by $\sim \pi^2/[(ak_{\rm F})^2|\d|]$ into ${\rm Im}\int
g_\br^2/(V^2 D) d\br/a^2$ in Eq. \ref{eq14}. Thus we arrive at the
estimate: $\gamma \sim c^2 V/|\d|$, and hence to the finite residual
DOS:

\be
 \r(\e_0) \sim  \r_N\frac{c^2 V}{\D|\d|}  \ln \frac {\D|\d|}{c^2 V}.
 \lb{eq16}
  \ee

We recall that this result is impossible in the properly formulated
self-consistent approximation \cite{lp}, and it is also in a
striking difference to the SCTMA predictions, $\r(0)\sim (\D/ V^2
\rho_N) \exp(-\D/c V^2\rho_N)$ in the Born limit \cite{gork} or
$\r(0)\sim\sqrt{c\rho_N/\D}$ in the unitary limit \cite{lee}. The
non-universality of this effect is manifested by its sensitivity to
the M-perturbation parameter $V$, so that $\r(\e_0)$ is defined by
impurity pairs only for strong enough perturbations, $|V| \gtrsim
(\r_N + g_{as})^{-1}$. For weaker $V$, nodal quasiparticles will be
only localized on impurity clusters of a greater number $n > 2$ (the
bigger the smaller $|V|$), and the particular form of Eq. \ref{eq16}
would change to $\rho(0)\sim \rho_N c^n$ (with some logarithmic
corrections). Anyhow, a finite limit of DOS at zero energy is
granted by the fact that in real high-\emph{T}$_{c}$ systems a
certain M-type perturbation can result even from nominally
non-magnetic centers (including dopants) \cite{lp3}. Moreover, the
above considered condition $|V|g_{as} \approx 1$ does not seem very
difficult, as testified by the observation of extremely low-energy
resonance $\e_{res} \approx -1.5$ meV by Zn impurities in
Bi$_2$Sr$_2$CaCu$_2$O$_{8+\d}$ \cite{pan}. It just fits the
asymmetric M-resonance from Fig. \ref{fig5}, contrasting with the
symmetric NM-resonance picture.

Then the overall impurity effect in a \emph{d}-wave superconductor
can be seen as a superposition (almost independent) of the above
described effects from NM-impurities with perturbation parameter
$V_{\rm NM}$ and concentration $c_{\rm NM}$ and from M-impurities
with perturbation parameter $V$ and concentration $c$ (supposedly $c
\ll c_{\rm NM}$). An example of such situation is shown in Fig.
\ref{fig5}.

\begin{figure}
\center\includegraphics[width=8.5cm]{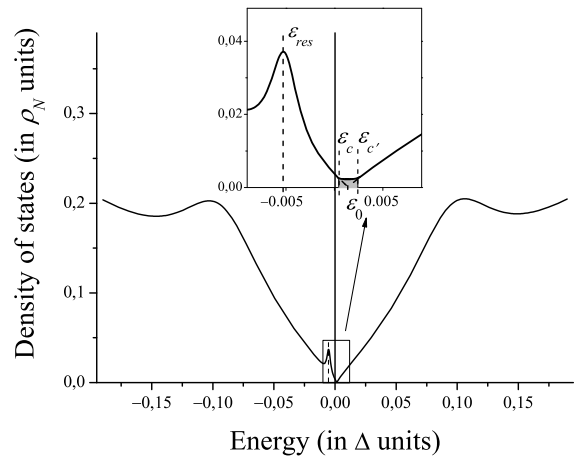}
 \caption{Low-energy \emph{d}-wave DOS $\r(\e)$ in simultaneous
 presence of NM-impurities (with $c_{\rm NM} = 3\%$ and $V_{\rm NM}
 = 1/\r_N$), producing two symmetric broad resonances, and M-impurities
 (with $c = 0.03 \%$, $V = V_{\rm NM}$, and $V g_{as}  = 0.9$),
 producing single sharp resonance at extremely low energy $\e_{res}$.
 Inset shows the mobility edges $\e_c$ and $\e_{c^\prime}$ around the
 shifted nodal point $\e_0$, they separate localized states (shadowed area)
 with almost constant DOS, $\r(\e) \approx \r(\e_0)$, from band-like states
 whose DOS is close to the T-matrix value (solid line).}
 \lb{fig5}
\end{figure}

The residual DOS from pair GE term prevails within a certain narrow
vicinity of $\e_0$ where quasiparticle states are all localized on
properly separated impurity pairs. Outside this vicinity, the states
are extended and reasonably described by T-matrix (or SCTMA). The
transition from localized to extended states occurs at the Mott
mobility edges $\e_c < \e_0$ and $\e_{c^\prime} > \e_0$, where GE
and T-matrix contributions to DOS are comparable. Using the simplest
approximation for the T-matrix term: $\r(\e) \sim \r_N|\e -
\e_0|/\D$, we estimate the range of localized states (somewhat
exaggerated in the inset of Fig. \ref{fig5}) as

\be \e_{c^\prime} - \e_0 \sim \e_0 - \e_c \sim \d_c \sim \frac{c^2
V}{|\d|}\ln\frac{\D|\d|}{c^2 V}, \lb{eq17} \ee

\noindent provided it is much smaller than the distance to the
M-resonance: $\d_c \ll |\e_0 - \e_{res}|$. The same estimates for
the mobility edges follow from the IRM breakdown condition: $\e -
{\rm Re}\S(\e) \sim {\rm Im}\S(\e)$, at $\e \approx \e_0$.

The tendency to localization of quasiparticles can be generally
opposed by the effects of repulsive Coulomb interaction between them
\cite{bas} and this issue was also discussed for disordered
\emph{d}-wave superconductors \cite{khve,fabr}. These field theory
treatments showed that localization can survive at low enough
temperature. The full account of Coulomb interactions in the present
GE approach is rather complicated technically, but a simple estimate
follows from the overall number of (supposedly) localized particles
within the energy range, Eq. \ref{eq17}, which is as small as
$n_{loc} \sim \r(\e_0) \d_c \sim (c^4 V/\D\d^2)\ln^2|\D\d/c^2V|$.
Since the average distance between them $\sim a/\sqrt{n_{loc}}$ is
much longer than the distance between charge carriers $\sim a/\sqrt
n \sim a/\sqrt{\r_N\e_{\rm F}}$, the effects of Coulomb interaction
are hopefully screened out, at least for the systems far enough from
half-filling \cite{boe}.

Notably, localization turns to be yet possible near the resonance
energy, $\e \approx \e_{res}$, but this requires that the
concentration of M-impurities surpasses a certain characteristic
value $c_{res} \sim (\e_{res}/\D)^2 \ln( \D/ |\e_{res}|)$. In
particular, for the choice of parameters in Fig. \ref{fig5}, we find
$c_{res} \sim 3\cdot 10^{-4}$, so that this system should be close
to the onset of localization also in this spectrum range, where each
localized state is associated with a \emph{single} impurity center.

Generally, presence of localized states near the lowest excitation
energies in the spectrum must influence significantly the kinetic
properties of a crystal with impurities, such as electric and
(electronic part of) heat conductivity at lowest temperatures.
Taking in mind the above referred modification of Kubo formula for
the energy range $|\e-\e_0| < \d_c$, their temperature dependencies,
instead of reaching the universal values $\sigma_{0}$ and
$[\kappa/T](0)$, should rather tend to the exponential vanishing:
$\sim\exp\left(-\d_c/k_{{\rm B}}T\right)$, at low enough
temperatures: $T \ll \d_c /k_{{\rm B}}$. The latter value, at the
same choice of impurity perturbation parameters and typical gap $\D
\sim 30$ meV, is estimated as $\sim 0.2$ K. In this context, the
intriguing sharp downturn of $\kappa/T$, recently observed at
temperatures $\lesssim 0.3$ K \cite{hill} and attributed to the
low-temperature decoupling of phonon heat channel \cite{smith}, can
be otherwise considered as a possible experimental manifestation of
the quasiparticle localization by impurity clusters. A more detailed
analysis of possible non-universal behavior of transport properties
of disordered \emph{d}-wave superconductors will be necessary to
confirm this conjecture.

\section{Conclusions}
The quasiparticle states are considered in a \emph{d}-wave
superconductor with impurities beyond the (self-consistent) T-matrix
approximation, using the techniques of group expansions of Green
functions in complexes of interacting impurities. It is shown that,
if the impurity perturbation of magnetic type is present, the
indirect interaction between impurities can essentially change the
quasiparticle spectrum near nodal points, producing strongly
localized states of non-universal character (depending on the
perturbation strength). Experimental check for possible
non-universal effects in low temperature transport properties can be
done, e.g., in the Zn doped Bi$_2$Sr$_2$CaCu$_2$O$_{8+\d}$ system.

\end{document}